\documentclass[twocolumn]{aastex63}

\newcommand{\spNyq}{f_{\rm SP}}
\newcommand{\ptaNyq}{f_{\rm PTA}}
\newcommand{\Np}{N_{\rm p}}

\usepackage{graphicx}
\usepackage{amsmath}
\usepackage{amssymb}
\usepackage{color}
\usepackage{float}
\usepackage{natbib}

\shorttitle{Extending the frequency reach of PTA based GW search}
\shortauthors{Wang, Mohanty, and Cao}

\begin{document}

\title{Extending the frequency reach of pulsar timing array based gravitational wave search \\ 
without high cadence observations}

\correspondingauthor{Y. Wang, S. D. Mohanty, Z. Cao}
\email{ywang12@hust.edu.cn, soumya.mohanty@utrgv.edu, zjcao@bnu.edu.cn}

\author[0000-0001-8990-5700]{Yan Wang}
\affiliation{MOE Key Laboratory of Fundamental Physical Quantities Measurements,
Hubei Key Laboratory of Gravitation and Quantum Physics, PGMF, Department of Astronomy and
School of Physics, Huazhong University of Science and Technology,
Wuhan 430074, China}

\author[0000-0002-4651-6438]{Soumya D. Mohanty}
\affiliation{Department of Physics and Astronomy, The University of Texas Rio Grande Valley, \\
One West University Boulevard, Brownsville, TX 78520, USA}

\author[0000-0002-1932-7295]{Zhoujian Cao}
\affiliation{Department of Astronomy, Beijing Normal University, Beijing 100875, China}

\begin{abstract}

Gravitational wave (GW) searches using pulsar timing arrays (PTAs) are  
assumed to be limited by the typical average  observational cadence of 
$1/(2~{\rm weeks})$ for a single pulsar to GW frequencies  $\lesssim 4\times 10^{-7}$~Hz.
We show that this assumption is incorrect and that a PTA can detect signals 
with much higher frequencies, which are preserved in the data due to aliasing, 
by exploiting asynchronous observations from multiple pulsars. 
This allows an observation strategy that is scalable to future large-scale PTAs containing
$O(10^3)$ pulsars, enabled by the Five-hundred meter 
Aperture Spherical Telescope and the Square Kilometer Array, 
without requiring a higher per-pulsar observation cadence. We show that 
higher frequency GW observations, reaching up to $4\times 10^{-4}$~Hz  
with an SKA-era PTA, have significant astrophysical implications, such as (i)
a three orders of magnitude better constraint than current high-cadence observations 
on GW strain in the $[10,400]$~$\mu{\rm Hz}$ band, and (ii) sensitive tests of the no-hair 
theorem in the mass range of supermassive black hole binaries
using their inspiral, merger, and ringdown signals. 

\end{abstract}


\section{Introduction} \label{sec:intro}

The ever growing trove of gravitational wave (GW) signals from compact binary 
coalescences~\citep{2016PhRvL.116f1102A, 2019PhRvX...9c1040A} collected by 
the LIGO~\citep{2015CQGra..32g4001L} and  Virgo~\citep{2015CQGra..32b4001A} 
detectors is revealing the GW universe in the $\sim 10$~Hz to $10^3$~Hz band. 
At lower frequencies, the space-based LISA~\citep{2017arXiv170200786A} mission 
will target the millihertz band from $10^{-4}$~Hz to $10^{-1}$~Hz while pulsar timing 
arrays (PTAs) are already putting meaningful constraints in the sub-$\mu$Hz band on
the stochastic GW background from an unresolved 
population of supermassive black hole binaries (SMBHBs) 
\citep{eptastochastic2015,2015Sci...349.1522S,2016ApJ...821...13A,2018ApJ...859...47A}, 
continuous waves from resolvable SMBHBs 
\citep{2014MNRAS.444.3709Z,2016MNRAS.455.1665B, 2019ApJ...880..116A}, 
and bursts \citep{2015MNRAS.446.1657W, 2020ApJ...889...38A}. 

The numbers of millisecond pulsars currently being timed by PTA consortia are 47 \citep[NANOGrav]{Alam_2020}, 
26 \citep[PPTA]{2020PASA...37...20K}, 42 \citep[EPTA]{2016MNRAS.458.3341D}, and 65 \citep[IPTA]{2019MNRAS.490.4666P}.
Next-generation radio telescopes, namely, the Five-hundred meter Aperture Spherical Telescope (FAST)~\citep{2011IJMPD..20..989N,2019RAA....19...20H} and the Square Kilometer Array (SKA)~\citep{2009A&A...493.1161S,2015aska.confE..37J} will grow the number of well-timed pulsars (noise rms $\lesssim 100$~ns) to $O(10^3)$. Along with a more uniform sky coverage and standardized data spans, this will improve the sensitivity to GWs from resolvable sources by two orders of magnitude~\citep{2017PhRvL.118o1104W,Erratum2017}.

The high frequency limit of the sensitive band for PTA based GW searches is widely assumed \citep{eptastochastic2015,2015Sci...349.1522S,2016ApJ...821...13A,2018ApJ...859...47A,2014MNRAS.444.3709Z,2016MNRAS.455.1665B, 2019ApJ...880..116A}  to be $\approx 4 \times 10^{-7}$ Hz, corresponding to the
Nyquist rate~\citep{2000fta..book.....B} associated with the average cadence 
of timing observations, typically $1/(2~{\rm weeks})$, for individual pulsars in an array. 
Therefore, attempts at extending the high frequency limit for resolvable GW sources, 
to frequency $> 1~\mu{\rm Hz}$, are all based on high cadence observations of single pulsars 
\citep{2010MNRAS.407..669Y,2014MNRAS.445.1245Y,2018MNRAS.478..218P,Dolch_2016}. 
It should be noted here that the actual cadences for pulsars in current PTAs vary over a large range and 1/(2 weeks) is more representative of its higher end.

In this letter, we show that the high frequency reach of PTAs for resolvable sources is 
not limited by the Nyquist rate, $\spNyq$, of single pulsar observations and that the
limiting frequency, $\ptaNyq$, can be much higher than assumed so far. 
The key here is that a higher frequency signal  is
preserved due to aliasing in the sequence of timing observations from each pulsar and can be unscrambled 
using asynchronous observations \citep{Wong2006stg,Bretthorst2001} from multiple pulsars. 

The lack of synchronicity, an inherent feature of PTA data, can be turned into an observational strategy, 
which we call {\it staggered sampling}, to boost the high frequency reach of PTAs. 
Staggered sampling simply requires the introduction, by design, of relative time shifts between 
the sequences of timing observations without requiring a change in the individual observational 
cadence for any pulsar. Unlike high cadence observations, this approach is scalable to future 
large-scale PTAs with $O(10^3)$ pulsars since it does not increase the total 
telescope time consumed by PTA observations. 

While the sensitivity of PTA-based GW searches falls with 
increase in GW signal frequency, an increase in the number of pulsars enhances it. 
This motivates a first exploration in this letter of the
astrophysical implications of high frequency searches with an SKA-era PTA, where staggered sampling could
increase the frequency reach to $\approx 4\times 10^{-4}$~Hz and
bridge the gap in coverage of the GW spectrum between PTAs and LISA.

\section{Preliminaries} \label{sec:pre}

In a PTA with $\Np$ pulsars, the timing residual of 
the $I$-th pulsar after subtracting a best-fit  
model (excluding GWs) of the pulse time of arrival 
is given by $d^I(t) =  s^I(t) + n^I(t)$, where $s^I(t)$ is the GW induced signal 
and $n^I(t)$ is noise. 

At the high signal frequencies of interest to us, the samples of
$n^I(t)$ can be assumed to be drawn from an
independent and identically distributed Normal random process 
with zero mean and constant variance $(\sigma^I)^2$ 
(i.e., white Gaussian noise). 
The contribution to $n^I(t)$ from errors in 
fitting the timing model are negligible at higher frequencies 
except at very specific ones \citep{2014PhRvD..89d2003C, 2004MNRAS.355..395K} 
such as $1$~${\rm yr}^{-1}$ and harmonics. The latter 
are ignored in our analysis due to the extremely narrow bands that are affected. 

With $\nu^I(t)$ and $\nu^I_0(t)$  denoting
the pulsar rotation frequencies observed at the Solar System Barycenter 
and at the pulsar, respectively, $s^I(t)$ is given by \citep{1975GReGr...6..439E, 2010PhRvD..81j4008S} 
\begin{equation}\label{eq:signal}
s^I(t) = \int_{0}^{t} \text{d}t' z^{I}(t')   \,,
\end{equation}
where  $z^I(t)\equiv (\nu^I(t)-\nu^I_0(t))/\nu^I_0(t)$  is the GW induced Doppler shift. 
For a plane GW arriving from right ascension ($\alpha$) and 
declination ($\delta$), with polarizations $h_{+,\times}(t;\theta)$ parametrized by source parameters $\theta$, 
\begin{equation}\label{eq:redshift}
z^I(t)  =  \sum_{A=+,\times} F^{I}_{A}(\alpha,\delta) \Delta h_{A} (t;\theta)   \,,
\end{equation}
\begin{equation}\label{eq:hpc}
\Delta h_{+,\times} (t;\theta)  =  h_{+,\times}(t;\theta)  - 
h_{+,\times}(t - \kappa^I;\theta)  \,,
\end{equation}
where, $F^{I}_{+,\times}(\alpha,\delta)$ are the antenna pattern functions \citep{2011MNRAS.414.3251L} 
and $\Delta h_{+,\times}$, for the two-pulse response \citep{1975GReGr...6..439E}, 
contain the so-called Earth and pulsar terms that arise from the action of 
the GW on pulses at the time, $t$, of their reception and at the time, $t-\kappa^I$, 
of their emission, respectively. 

For a non-evolving circular binary emitting a monochromatic signal,  $\theta$ includes  
the overall amplitude ($\zeta$), GW frequency ($f_{\rm gw}$), 
inclination angle of the binary orbital angular momentum relative to the line of sight ($\iota$), 
GW polarization angle ($\psi$), and initial orbital phase ($\varphi_0$) \citep{2014ApJ...795...96W}. 
The time delay $\kappa^I$ appears as an unknown constant phase offset, 
called the pulsar phase parameter $\phi_I$, for such a source. 

To search for resolvable GW sources, we 
use the likelihood based detection and parameter estimation method 
described in \citep{2015ApJ...815..125W,Wang_2017,RAAPTR} 
that takes both the Earth and pulsar terms into account. 
The method partitions the estimation of parameters such that the pulsar 
phases are either maximized \citep{2015ApJ...815..125W} or, as chosen here, marginalized \citep{Wang_2017} semi-analytically, 
allowing the method to scale to an arbitrarily large $\Np$.  The 
remaining parameters are estimated numerically by maximizing the (marginalized)
likelihood using Particle Swarm Optimization~\citep{PSO,mohanty2018swarm,2016MNRAS.461.1317Z}.  
The maximum value serves as the detection statistic for deciding between the
null ($H_0$) and alternative ($H_1$) hypotheses about given data that a signal is absent or present, 
respectively.

\section{Staggered sampling} \label{sec:stg}

Let $t^I = \{t_i^I\}$, $i = 1, 2, \ldots, N^I$, denote the times  at which
the residual $d^I(t)$ is sampled and let
their spacing, $t^I_{i+1}-t^I_i$, be $\Delta>0$ on the average 
(e.g., $\Delta \geq 2$~weeks for IPTA pulsars \citep{2016MNRAS.458.1267V}). 
Consider the set $\mathcal{T}=\cup_{I=1}^{\Np}t^I$ of sample times in ascending 
order from all the array pulsars and let $x_k$, $k=1,2,\ldots,\sum_{I=1}^{\Np} N^I$, 
denote an element of this set. 
We consider two specific schemes for staggered sampling in our 
analysis. In both of them, we set $N^I = N$ to be the same for all the pulsars.
This is mainly for reducing the complexity of our codes and not an essential 
requirement for staggered sampling. 

The most straightforward scheme, called {\it uniform staggered sampling}, 
 is one where $x_{i+1}-x_i$ is a constant. This implies that 
 the samples of $d^I(t)$ are uniformly spaced and the sequence of samples 
 from one pulsar has a constant time shift relative to those of others: 
 $t^{I}_{i} = (i-1) \Delta+\delta^I$, with $\delta^I = (I-1)\Delta/\Np$. 
 
 In the second scheme, called 
 {\it randomized staggered sampling}, $x_{i+1}-x_i$ is a random variable.
 This is a more realistic situation 
 given the uncertainties inherent in planning astronomical observations. 
 However, as with current PTAs, any real observation strategy would
 have a target that it seeks to approximate, which is assumed here to be uniform staggered sampling. 
 Therefore, we adopt a reasonable model for randomized staggered sampling 
 in which $t^{I}_{i}$ is replaced by $t^{I}_{i} + c^{I}_{i}$, where $c^{I}_{i}$ is a random variable. 
 In our analysis, we will assume that $c^{I}_{i}$ is drawn from a truncated
 Cauchy probability density function (pdf) \citep{1984prvs.book.....P}
 with a location parameter set to zero, scale factor of 
$1/3$ day, and $|c^{I}_{i}|\leq 7$~days. 
The heavy tails of this pdf allow large excursions -- the $99\%$ inter-percentile range is $\simeq 10$~days -- from the planned
observation times of uniform staggered sampling. 

The search for individual GW sources is carried out on
staggered sampling data as described earlier -- no changes are 
required to the detection and estimation algorithm as it works 
entirely in the time domain. That this leads to a higher frequency 
reach is validated directly in this letter
using simulated data. While a rigorous analytic treatment of
an arbitrary staggered sampling scheme is left to future work, 
the following argument indicates what the maximum detectable signal frequency 
should be. 
Take the trivial case of identical GW induced residuals, $s^I(t)=s^J(t)$, observed with uniform
staggered sampling.  Since the identical residuals lead to a 
common signal, pooling all data into a single time series will yield
samples of the same signal but with a smaller spacing of 
$\Delta/\Np$. Hence, the maximum detectable frequency is $\ptaNyq=\Np \spNyq$. 
Simply pooling the data does not work for the real case of a heterogeneous, $s^I(t) \neq s^J(t)$ 
for $I\neq J$, set of GW induced residuals but one expects the same limit to hold. 

Note that a higher cadence observational strategy to achieve the same high frequency 
limit as staggered sampling, namely $\Np\spNyq$, would increase the 
total telescope time occupied in timing observations by a factor $\Np$ since 
each pulsar must be timed with a cadence of $\Delta/\Np$. This makes 
the high cadence strategy  extremely costly and unviable for the large $\Np$ of $O(10^3)$ in an SKA-era PTA.

\section{Detection and parameter estimation} \label{sec:det}

We use the following simulation setup to show that the 
staggered sampling schemes described above increase the frequency reach of a PTA.
We consider a PTA with $\Np = 50$ nearest pulsars chosen from the simulated catalog in \citet{2009A&A...493.1161S}. 
The total observation period is set at $T=5$~yr with observations
spaced $\Delta = 2$~weeks apart in the case of uniform staggered sampling: this results in 
$\spNyq \approx 4 \times 10^{-7}~{\rm Hz}$. 
($T$ is set lower than the typical value of 10 yr or more for PTA data to keep computational costs of the simulation in check.) 
We consider non-evolving sources with 
four angular frequencies $\omega_{\rm gw}$: 
64 rad/yr ($3.23 \times 10^{-7}~{\rm Hz}$), 256 rad/yr ($1.29 \times 10^{-6}~{\rm Hz}$), 
1024 rad/yr ($5.16 \times 10^{-6}~{\rm Hz}$) 
and 4096 rad/yr ($2.0656 \times 10^{-5}~{\rm Hz}$). 
Note that the last three sources have frequencies $ > \spNyq$ and
the highest one is very close to the staggered sampling
limit of $\ptaNyq = \Np\spNyq = 4098.09$~rad/yr.  
The sources are located at $\alpha = 3.5$~rad and $\delta = 0.3$~rad in 
equatorial coordinates. 
This location corresponds to the lowest degree of ill-posedness in parameter estimation for the SKA-era PTA used in \cite{2017PhRvL.118o1104W}. 
The inclination angle 
and the GW polarization angle are given by $\iota = 0.5$~rad and $\psi =0.5$~rad, respectively. 
The initial orbital phase is set at $\varphi_0 = 2.89$~rad. 
 
Following the noise model described earlier, the 
standard deviation of the noise $n^I(t)$ is set at $\sigma^I = 100$~ns. The overall amplitude, $\zeta$,
of the GW signal, which 
depends on the distance to the source, its chirp mass, and GW frequency, 
is determined by the specified network signal-to-noise ratio (SNR) $\rho$. 
Here $\rho^2 = \sum_{I=1}^{\Np} (\rho^{I})^2$, where 
\begin{equation}\label{eq:snri}
(\rho^{I})^2  =  \frac{1}{(\sigma^{I})^2 \Delta }\int_0^T (s^I(t))^2 dt  
\end{equation}
is the squared SNR of the GW induced timing residual for 
the $I$-th pulsar.

\begin{figure}
\centerline{\includegraphics[width=0.53 \textwidth]{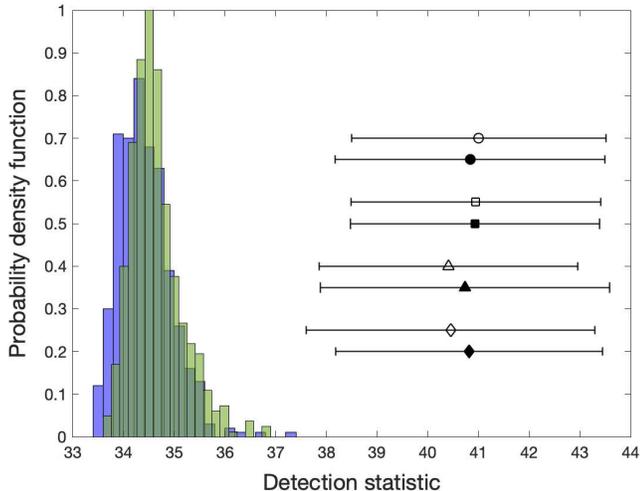}}
\caption{Distributions of the detection statistic under the $H_0$ (signal absent) and
$H_1$ (signal present) hypotheses. 
$H_0$: histograms obtained from $500$ data realizations are shown
for uniform (blue) and randomized (green) staggered sampling. 
$H_1$: the estimated mean (marker) and $\pm 1\sigma$ deviation (error bar)
of the detection statistic, obtained from 
$200$ data realizations, are shown for different signal angular frequencies,  
$\omega_{\rm{gw}}=64$ (circle), $256$ (square), $1024$ (triangle) 
and $4096$ (diamond) rad/yr, and uniform (open markers) or randomized (filled markers)
staggered sampling. In all cases, the signal SNR is $\rho = 10$. 
The vertical offset of an error bar or marker is for visual clarity only. } 
\label{fig:detection}
\end{figure}

Fig.~\ref{fig:detection} shows the distributions of
the detection statistic under the $H_0$ and $H_1$
hypotheses for both 
the uniform and the randomized staggered sampling 
strategies. From these, we estimate the detection 
probability of a $\rho = 10$ signal to be $\gtrsim 90 \%$ 
at a false alarm probability of $\simeq 1/500$. The latter
corresponds to setting 
the detection threshold at the largest value of the detection statistic
obtained from $H_0$ data realizations.
Within the precision of our simulation, these numbers
are fairly independent of the staggered sampling strategy and the 
GW signal frequency. While a two-sample Kolmogorov-Smirnoff test on the 
$H_0$ distributions
does show their apparent relative shift to be statistically significant, this has 
no noticeable effect on detection probability at the above SNR. 

Given that a higher frequency signal has a larger number of cycles in
a given observation period, it is natural to ask if it can be detected 
over shorter observation periods using staggered sampling. We verified this by repeating 
the above simulations with $T = 1$~yr and $\omega_{\rm gw} = 512$~rad/yr and $1024$~rad/yr
keeping all else fixed. 
The detection probabilities had insignificant changes,
suggesting that the performance of a staggered sampling based search depends 
primarily on the SNR of a signal. 

\begin{figure}
\centerline{\includegraphics[scale=0.45]{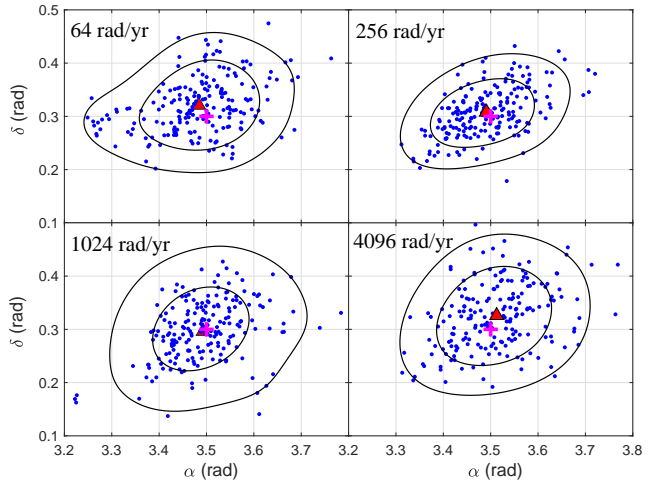}}
\caption{Distribution of sky location  in equatorial coordinates ($\alpha, \delta$) for signals with different angular frequencies using uniform staggered sampling. Each panel shows estimated sky locations (dots)  from $200$ data realizations, each containing an SNR $\rho = 20$ signal with an angular frequency, $\omega_{\rm gw}$, as noted in the panel.
The true location of the GW source
is  marked by a triangle and 
the mean of the estimated locations is marked by a cross. 
The solid contour lines are obtained using 
2D Kernel Density Estimation \citep{botev2010} and show regions with
areas $\Delta\Omega_{68\%}$ and $\Delta\Omega_{95\%}$ in which the probabilities of getting
estimated locations are 68\% and 95\%, respectively. 
For ascending $\omega_{\rm gw}$, $\Delta\Omega_{68\%}$($\Delta\Omega_{95\%}$) is  $103$($262$), $75$($186$), $77$($305$), and $115$($306$) 
${\rm deg}^2$. } 
\label{fig:contour2}
\end{figure}

Fig.~\ref{fig:contour2} shows the estimated sky locations of the source
for uniform staggered sampling ($T=5$~yr) and  a  moderately strong SNR of $\rho = 20$. 
Within the precision of our simulations, the error in localizing a GW source 
does not show a clear trend with the frequency of the signal. Resolving a trend, if it exists, would
require a computationally much more expensive simulation that we
leave for future work. 
The typical localization error in Fig.~\ref{fig:contour2} of 
$O(100)$~${\rm deg}^2$ makes searches for optical counterparts of GW sources 
feasible with the Rubin observatory \citep{2019ApJ...873..111I} 
across the entire range, $[T^{-1},\ptaNyq]$, of GW frequencies \citep{2041-8205-803-2-L16,2017PhRvL.118o1104W}. 

\section{Astrophysical implications} \label{sec:astro}

For an SKA-era PTA with $\Np = 10^3$ pulsars and per-pulsar 
observational cadence of $1/(2~{\rm weeks})$, our results show that 
staggered sampling can increase $\ptaNyq = \Np \spNyq$ to as high as $4\times 10^{-4}$ Hz. 
To estimate the achievable sensitivity, we use the same simulated PTA as in \citet{2017PhRvL.118o1104W}, 
comprised of millisecond pulsars within 3 kpc taken 
from the synthetic catalog in \citet{2009A&A...493.1161S}. 
From an analysis similar to Fig.~\ref{fig:detection}, 
we find that the detection probability of a monochromatic signal for the SKA-era PTA is $\simeq 60\%$ at $\rho = 10$ 
for a false alarm probability of $\simeq 1/50$.  We adopt this as the fiducial value for 
the minimum detectable SNR averaged over the 
sky angles $\alpha$ and $\delta$. (The resulting geometrical factor is $\simeq 1$ for this PTA.) 
Non-detection of a signal at this SNR will result in a sky-averaged upper limit 
on monochromatic GW strain amplitude, 
\begin{equation}\label{eq:rho2h}
h = 8.89 \times 10^{-15} \left(\frac{f_{\text{gw}}}{10^{-6}~\text{Hz}}\right) \left(\frac{T}{5~\text{yr}}\right)^{-\frac{1}{2}}\left(\frac{\sigma}{100~\text{ns}}\right)   \,, 
\end{equation}
that is about three orders of magnitude lower in the $[10,400]~\mu{\rm Hz}$ band than 
the current one from high cadence observation 
of millisecond pulsar J1713+0747 \citep{Dolch_2016}.

In the extended frequency range, 
not only would the inspiral phase of an SMBHB signal
be observable but also the merger and ringdown phases. 
To quantify the sensitivity to each of these phases, we use the luminosity
distance, $D_L$, for a sky-averaged ${\rm SNR}=10$. 
(The inclination and polarization angles are also averaged over in the case of inspirals). 
Since searches for the pulsar and Earth term can be decoupled for
a strongly evolving signal \citep{2010ApJ...718.1400F}, we make
the conservative choice of using the SNR of only the Earth term.

The inspiral signal is calculated in the Newtonian approximation \citep{1963PhRv..131..435P} 
over an observation period $\min\{20 \,{\rm yr}, \tau\}$,
where $\tau$ is the lifetime for the signal frequency to evolve from 
an initial value $f_i$ to $f_{\rm ISCO}$, the frequency at
the innermost stable circular orbit (ISCO). 
The merger and ringdown phases are obtained from waveforms 
computed in the spin aligned effective one body numerical relativity for eccentric 
binary (SEOBNRE) \citep{PhysRevD.96.044028,PhysRevD.101.044049} formalism: 
the part of the waveform
between the instantaneous frequency exceeding $f_{\rm ISCO}$ and the
instantaneous amplitude attaining its maximum value is defined as the merger, with
the subsequent phase being the ringdown. For the latter, only the dominant
$l=2, |m|=2$  mode, with its corresponding frequency $f_{2,2}$, is used. 
We consider only circular binaries with zero spin and equal mass components
for which the defining parameters are only the chirp mass  $\mathcal{M}_c = 0.435 \, M$ ($M$ is the total mass) and, 
for the inspiral, the chosen $\tau$.

\begin{figure}
\centerline{\includegraphics[scale=0.47]{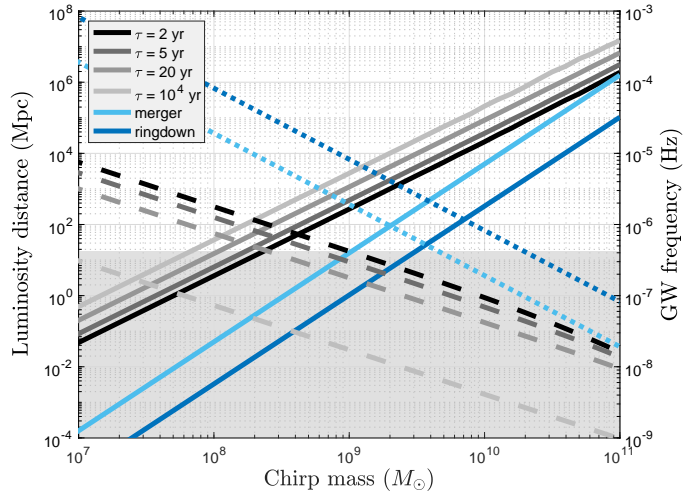}}
\caption{Luminosity distance, $D_L$,
(left $y$-axis and solid lines) and GW frequency (right $y$-axis and broken lines) 
as a function of chirp mass $\mathcal{M}_c$ for a maximum observation duration of $20$~yr. 
The gray shaded area covers $f \in [10^{-9},~\spNyq= 4\times 10^{-7}]$~Hz. 
Lines in gray scale represent inspirals with different lifetimes as 
indicated in the legend. 
Light and dark blue lines represent merger and ringdown, respectively. 
The frequencies shown are $f_i$, $f_{\rm ISCO}$, and $f_{2,2}$. 
}
\label{fig:DlFreq}
\end{figure}

Fig.~\ref{fig:DlFreq} shows $D_L$ for all the different phases 
as a function of $\mathcal{M}_c$ and $\tau$ along with their  characteristic frequencies. 
We see that for  $\mathcal{M}_{c}\geq 4.5\times 10^9$~$M_\odot$, the inspiral signal 
always stays below $\spNyq$ irrespective of $\tau$.  
On the other hand, the inspiral signal for 
$\mathcal{M}_{c} < 4.5\times 10^9$~$M_\odot$ would cross $\spNyq$ even if $f_i < \spNyq$. 
Table~\ref{tab:astro_results} summarizes the distance reach, with the distance ($D_L = 20$~Mpc) 
to the Virgo cluster as a baseline, for different 
signal phases that require the extended frequency range ($\geq \spNyq$)
of staggered sampling to be observable. Further applications of Fig.~\ref{fig:DlFreq} 
are considered below. 

\begin{table*}
    \centering
    \begin{tabular}{|c|c|c|c|c|c|c|}
    \hline
    Signal & \multicolumn{2}{c}{$D_L>20$~Mpc} & \multicolumn{2}{|c}{$D_L>100$~Mpc} 
    & \multicolumn{2}{|c|}{$D_L>500$~Mpc}\\ 
    \hline
    Inspiral   &   $[2.2, 45] \times 10^8 \;M_\odot$  &  $[2, 20]\; {\rm yr}$  
                  &   $[6, 45] \times 10^8 \;M_\odot$  &  $[2, 20]\; {\rm yr}$
                  &   $[15, 45] \times 10^8 \;M_\odot$  &  $[2, 20]\; {\rm yr}$ \\ 
                  \hline
    Merger   &  $[4, 45] \times 10^{8} M_{\odot}$  & $[3.7, 42] \; {\rm d}$
                  &  $[20, 45] \times 10^{8} M_{\odot}$  & $[19, 42] \; {\rm d}$   
                  &  $[37, 45] \times 10^{8} M_{\odot}$  & $[35, 42] \; {\rm d}$    \\
                  \hline
    Ringdown  &  $[3, 20] \times 10^{9} M_{\odot}$  & $[8, 54] \; {\rm d}$
                      &  $[6, 20] \times 10^{9} M_{\odot}$  & $[16, 54] \; {\rm d}$   
                      &  $[13, 20] \times 10^{9} M_{\odot}$  & $[32, 54] \; {\rm d}$   \\
                  \hline
    \end{tabular}
    \caption{The range of chirp mass, $\mathcal{M}_{c}$, and 
    lifetime (or duration)  of different signal phases that lead to a given minimum luminosity distance, $D_L$, 
    under staggered sampling. 
In all cases above,  $f_{\rm ISCO}\geq \spNyq$ for inspiral and merger  while $f_{2,2} \geq \spNyq$ for ringdown signals. The minimum $D_L$ in each case corresponds to the respective lowest lifetime (or duration). 
The duration for a ringdown is taken as 3 times its exponential damping time scale. }
    \label{tab:astro_results}
\end{table*}

The observation of ringdown signals by an SKA-era PTA with staggered sampling 
could extend the test of the no-hair theorem to the extremely large mass
range of SMBHB remnants. 
For this we consider the test in \cite{PhysRevLett.123.111102} 
that achieves an $\approx 10\%$ level as defined by 
the fractional difference in the estimated values 
of a particular combination of mass and spin parameters measured from the late ringdown 
and the (${\rm SNR}\approx 14$) post peak-amplitude waveforms of GW150914.

For the above SNR of the post peak-amplitude (our ringdown) waveform, staggered sampling will allow the ringdown from 
a $\mathcal{M}_{c}\lesssim 2\times 10^{10}$~$M_\odot$ system, 
for which $f_{2,2}\gtrsim \spNyq$, to be detected out to $D_{L} \lesssim 1.32$~Gpc. 
Given that the corresponding inspiral signal 
would be extremely loud, ${\rm SNR}=620$ for $\tau> 2$~yr, 
the source would be localized well in advance of the ringdown. This would allow a
subset of favorably 
located pulsars to be targeted for significantly better timing over the 
duration of the ringdown. Assuming the 
timing residual noise is reduced from $100$~nsec 
to $\approx 20$~nsec \citep{PhysRevD.102.023014}, 
the observed ringdown SNR would increase to $\approx 70$, leading to a test of the
no-hair theorem at the $\approx 2\%$  level.

Considering a lower mass system such as $\mathcal{M}_{c}= 5\times 10^8$~$M_\odot$,
Fig.~\ref{fig:DlFreq} shows that a
$\tau = 5$~yr inspiral signal, with $f_i\gtrsim\spNyq$,
will be detectable out to $D_L \approx 100$~Mpc.
While the corresponding  ringdown 
signal ($f_{2,2}=0.02$~mHz) would be too weak for the SKA-era PTA, 
it would be extremely loud, with ${\rm SNR}\approx 220$, for a concurrently
operating LISA. 
Compared to the fiducial ringdown SNR above, 
the relative measurement accuracy of all the 
ringdown parameters -- inversely proportional to SNR from Fisher information 
analysis \citep{PhysRevD.73.064030} -- would improve by 
a factor of $\approx 16$. In combination with the PTA detected inspiral, this would again lead to a stringent test of the no-hair theorem.

In recent work \citep{2020arXiv200906084D}, a scheme for 
measuring the Hubble constant, $H_0$, has been proposed that uses only GW observations by an SKA-era PTA without requiring an electromagnetic counterpart. It relies on measuring
both $D_L$ and the 
co-moving distance $D_{c}= D_L/(1+z)$, $z$ is the 
cosmological redshift, of a GW source through the effect of GW wavefront curvature on 
pulsar timing residuals. The governing condition 
for the measurability of this effect is \citep{2020arXiv200906084D} $\gamma = 
\pi f_{\rm gw} L^2/D_c \gtrsim 0.1$, where $L$ is the Earth-pulsar distance.
For the assumed GW frequency $f_{\rm gw} =10^{-7}\,{\rm Hz} < \spNyq$ in \cite{2020arXiv200906084D}, 
achieving a precision of $\delta H_{0}/H_{0} \lesssim 30\%$ in this scheme puts a rather 
stringent observational requirement 
on the error, $\delta L$, in $L$ of $\delta L/L \sim 1\%$ at $L> 10$~kpc. 
However,  the feasibility of this scheme is improved if staggered sampling is used
to reach higher $f_{\rm gw}$. 
For example, $L$ reduces to $3$~kpc for the same relative error if  $f_{\rm gw}= 10^{-6}$~Hz. 
This could happen if $f_{\rm gw}=f_{\rm ISCO}$ for an $\mathcal{M}_{c}= 2\times 10^{9}$~$M_\odot$ system, which would be detectable out to  $D_L = 2$~Gpc ($4$~Gpc) for $\tau = 5$~yr ($ 20$~yr), yielding $\gamma = 0.21$ ($0.125$).

\section{Discussion} \label{sec:diss}

The impact of the extended frequency reach from staggered sampling on the 
detectability of a wider range of signals than considered here needs further study.
Among these are higher signal harmonics \citep{1963PhRv..131..435P} from unequal mass
SMBHBs that, orbital evolution studies indicate \citep{2010ApJ...719..851S}, 
could be driven to high eccentricities ($\sim 0.3$) by interactions with the stellar environment. 
Independent evidence comes from observations \citep{2018ApJ...866...11D}
of the SMBHB candidate OJ 287 that suggest  
a binary mass ratio of $\simeq 122$ and eccentricity $0.657$. 
Besides SMBHBs, oscillation of a network of 
cosmic strings \citep{2019A&ARv..27....5B}, superradiance from axion clouds around
isolated black holes \citep{PhysRevD.95.124056}, near-zone waves induced 
by turbulence of solar convection \citep{Bennett_2014}, and solar oscillation 
modes \citep{1996PhRvD..54.1287C, PhysRevD.79.082001} could be potential 
targets for a staggered sampling based search.

\acknowledgments

Y. W. gratefully acknowledges support from 
the National Natural Science Foundation of China (NSFC) 
under Grants No. 11973024 and No. 11690021, and 
Guangdong Major Project of Basic and Applied Basic Research (Grant No. 2019B030302001). 
The contribution of S. D. M. to this paper is supported by NSF 
Grant No. PHY-1505861. 
Z. C. gratefully acknowledges support from NSFC 
under Grant No. 11690023.  
We thank Wen-Fan Feng for discussions on LISA. 
We acknowledge the Texas Advanced Computing Center (TACC) at 
the University of Texas at Austin (www.tacc.utexas.edu) for providing high performance computing resources.
We thank the anonymous referee for helpful comments and suggestions.

\bibliographystyle{aasjournal}

\end{document}